\documentclass[12pt]{article}
\usepackage{setspace} 
\usepackage{indentfirst}
\usepackage{amsmath}
\usepackage{amssymb}
\usepackage{fullpage}
\usepackage{graphicx}

\usepackage{color}
\usepackage{natbib}
\usepackage{booktabs,amsmath}
\usepackage{sectsty}
\usepackage{yfonts}

\sectionfont{\small}
\subsectionfont{\small}

\title{\large The slow slip of viscous faults} 
\author{\small Robert C. Viesca\footnote{Dept. Civil and Environmental Engineering, Tufts University, 200 College Avenue, Medford, MA, USA}, Pierre Dublanchet\footnote{MINES ParisTech, PSL Research University, Centre de G\'eosciences, 35 rue Saint-Honor\'e, 77300 Fontainebleau, France}}
\date{\small December 4, 2018}

\begin{document}
\maketitle
\small

\begin{abstract}
We examine a simple mechanism for the spatio-temporal evolution of transient, slow slip. We consider the problem of slip on a fault that lies within an elastic continuum and whose strength is proportional to sliding rate. This rate dependence may correspond to a viscously deforming shear zone or the linearization of a non-linear, rate-dependent fault strength. We examine the response of such a fault to external forcing, such as local increases in shear stress or pore fluid pressure. We show that the slip and slip rate are governed by a type of diffusion equation, the solution of which is found using a Green's function approach. We derive the long-time, self-similar asymptotic expansion for slip or slip rate, which depend on both time $t$ and a similarity coordinate $\eta=x/t$, where $x$ denotes fault position. The similarity coordinate shows a departure from classical diffusion and is owed to the non-local nature of elastic interaction among points on an interface between elastic half-spaces. We demonstrate the solution and asymptotic analysis of several example problems. Following sudden impositions of loading, we show that slip rate ultimately decays as $1/t$ while spreading proportionally to $t$, implying both a logarithmic accumulation of displacement as well as a constant moment rate. We discuss the implication for models of post-seismic slip as well as spontaneously emerging slow slip events.
\end{abstract}


\

\section{Introduction}

A number of observations point towards the slow, stable slip of faults in the period intervening earthquakes. These include observations indicating slow slip following an earthquake---also known as post-seismic slip, or afterslip---and transient events of fault creep that appear to emerge spontaneously, without a preceding earthquake. Models for transient episodes of slow slip look to couple fault sliding with elastic or visco-elastic deformation of the host rock and incorporate various constitutive descriptions for the fault shear strength. We outline prior evidence for slow fault slip, a survey of past models for such behavior, and highlight an outstanding problem whose solution bridges existing gaps among models and between models and observations.


\subsection{Slow slip observations}

Early inference of fault creep, including creep transients and post-seismic slip, relied on the measurement of relative displacement of markers at the surface, designed instruments, and repeated geodetic surveys [e.g., \emph{Steinbrugge et al.}, 1960; \emph{Smith and Wyss, 1968}; \emph{Scholz et al.}, 1969; \emph{Allen et al.}, 1972; \emph{Bucknam}, 1978; \emph{Evans}, 1981; \emph{Beavan et al.}, 1984; \emph{Williams et al.}, 1988; \emph{Bilham}, 1989; \emph{Gladwin}, 1994; \emph{Linde et al.}, 1996]. However, using surface displacement measurements to discriminate between fault slip or more distributed deformation or using surface offset measurements to discriminate between shallow or deep sources of relative displacement were not typically possible, or at least attempted, owing to the paucity of information or computational resources.

\

The increased spatial-temporal resolution of satellite-based geodetic (chiefly, GPS and InSAR) led to more robust inference of aseismic fault slip. Among the earliest applications was for the inference of post-seismic slip. Typically, this inference was based on the goodness of fit of subsurface dislocation models to observed post-seismic surface displacement. Occasionally, in an attempt to discriminate the source of the post-seismic deformation, the goodness of such fits were compared to those with models that alternatively or additionally included mechanisms for distributed deformation, such as viscoelastic relaxation of the asthenosphere or poroelastic rebound of the crust. Notable earthquakes from which post-seismic slip has been inferred using such approaches include Landers '92 [\emph{Shen et al.}, 1994; \emph{Savage and Svarc}, 1997]; Japan Trench '94 [\emph{Heki et al.}, 1997]; Colima '95 [\emph{Az\'ua et al., 2002}]; Kamchatka '97 [\emph{Burgmann et al.}, 2001]; Izmit '99 [\emph{Reilinger et al.}, 2000; \emph{\c{C}akir et al.}, 2013]; Chi-Chi '99 [\emph{Yu et al.}, 2003; \emph{Hsu et al.}, 2002, 2007];  Denali '02 [\emph{Johnson et al.}, 2009]; Parkfield '04 [\emph{Murray and Segall}, 2005; \emph{Murray and Langbein}, 2006; \emph{Johnson et al.}, 2006; \emph{Freed}, 2007; \emph{Barbot et al.}, 2009; \emph{Bruhat et al.}, 2011]; Sumatra '04 [\emph{Hashiomoto et al.}, 2006; \emph{Paul et al.}, 2007]; Nias '05 [\emph{Hsu et al.}, 2006]; Pisco '07 [\emph{Perfettini et al.}, 2010]; Maule '10 [\emph{Bedford et al.}, 2013]; and Tohoku '11 [\emph{Ozawa et al.}, 2011, 2012].

\

Nearly in parallel, aseismic transients not linked to large earthquakes were discovered on subduction zones in Japan [\emph{Hirose et al.}, 1999; \emph{Hirose and Obara}, 2005; \emph{Obara and Hirose}, 2006; \emph{Hirose et al.}, 2014], Cascadia [\emph{Dragert et al.}, 2001; \emph{Miller et al.}, 2002; \emph{Rogers and Dragert}, 2003],  Guerrero, Mexico [\emph{Lowry et al.}, 2001; \emph{Kostoglodov et al.}, 2003], New Zealand [\emph{Douglas et al.}, 2005; \emph{Wallace and Beavan}, 2006], Alaska [\emph{Ohta et al.}, 2006; \emph{Fu and Freymueller}, 2013]; the Caribbean [\emph{Outerbridge et al.}, 2010], as well as several strike-slip faults [e.g., \emph{de Michele et al.}, 2011; \emph{Shirzaei and Burgmann}, 2013; \emph{Jolivet et al.}, 2013; \emph{Rousset et al.}, 2016]. In addition to the geodetic inference of slip, tremor was also observed seismologically, accompanying subduction zone slow slip events [\emph{Obara}, 2002; \emph{Rogers and Dragert}, 2003; \emph{Obara et al.}, 2004; \emph{Shelley et al.}, 2006; \emph{Obara and Hirose}, 2006; \emph{Ito et al.}, 2007]. The tremor is at least partly comprised of small, low frequency earthquakes with indications that these events occur as the rupture of small asperities driven by aseismic creep of the surrounding fault [\emph{Shelly et al.}, 2006; \emph{Shelly et al.}, 2007; \emph{Ide et al.}, 2007; \emph{Rubinstein et al.}, 2007; \emph{Bartlow et al.}, 2011]. Thus, the presence of tremor alone is a potential indication of underlying slow slip, which may be too small or deep to be well resolved geodetically, in both subduction and strike-slip settings [e.g., \emph{Nadeau and Dolenc}, 2005; \emph{Gomberg et al., 2008}; \emph{Shelly and Johnson}, 2011; \emph{Wech et al.}, 2012; \emph{Guilhem and Nadeau}, 2012]. 


\subsection{Models for slow slip}

Following initial observations of unsteady fault creep, modeling efforts looked towards continuum models to reproduce surface observations. Models represented fault slip as a dislocation within an elastic medium and transients as phenomena due to propagating rupture fronts or emerging from an explicit rate-dependent frictional strength of the fault [e.g., \emph{Savage}, 1971; \emph{Nason and Weertman}, 1973; \emph{Ida}, 1974; \emph{Wesson}, 1988]. However, because of the sparsity of available data with which to discriminate among hypothetical model assumptions, these representations may have been before their time.

\

An experimental basis for forward models followed laboratory rock friction experiments and the subsequent development of a rate-dependent or a rate- and state-dependent constitutive formulation for fault frictional strength [\emph{Dieterich}, 1978, 1979; \emph{Ruina} 1980, 1983]. Incorporation of the rate-and-state description into spring-block models  yielded a low-parameter model for both slow and fast fault slip defined in terms of frictional properties, a representative elastic stiffness, and a driving force [\emph{Rice and Ruina}, 1983; \emph{Gu et al.}, 1984; \emph{Rice and Tse}, 1986; \emph{Scholz}, 1990; \emph{Marone et al.}, 1991; \emph{Ranjith and Rice}, 1999; \emph{Perfettini and Avouac}, 2004; \emph{Helmstetter and Shaw}, 2009]. For instance, the analysis of such a single-degree-of-freedom model provided a simple representation of post-seismic relaxation of fault slip and simple scaling of displacement and its rate with time capable of matching observed displacement time history at points on the surface [e.g., \emph{Scholz}, 1990; \emph{Marone et al.}, 1991; \emph{Perfettini and Avouac}, 2004].

\

The increased availability of field observations led to a resurgence of continuum forward models, capable of matching multiple station observations with a single parameterized representation of the fault, as well as permitting the emergence of phenomena not possible within single-degree-of-freedom spring-block models. These include models for the afterslip process [e.g., \emph{Linker and Rice}, 1997; \emph{Hearn et al.}, 2002; \emph{Johnson et al.}, 2006; \emph{Perfettini and Ampuero, 2008}; \emph{Barbot et al.}, 2009; \emph{Hetland et al.}, 2010] as well as spontaneous aseismic transients [\emph{Liu and Rice}, 2005, 2007; \emph{Rubin}, 2008; \emph{Segall et al.}, 2010; \emph{Ando et al.}, 2010, 2012; \emph{Collela}, 2011; \emph{Skarbek et al.}, 2012; \emph{Shibazaki et al.}, 2012; \emph{Liu}, 2014; \emph{Li and Liu}, 2016; \emph{Romanet et al.}, 2018]. Such continuum models have been theoretically demonstrated to exhibit slowly propagating slip-pulse or detachment-front solutions, which provide additional mechanisms to mediate aseismic rupture, though  comparisons with observations are thus far limited [e.g., \emph{Perrin et al.}, 1995; \emph{Garagash}, 2012; \emph{Bar-Sinai et al.},  2012;  \emph{Putelat et al.}, 2016]. 

\

In both the continuum and spring-block slider models, rate- and state-dependent friction has remained as the prominent description of fault strength. The constitutive formulation consists of a positive logarithmic dependence on the sliding rate and an additional dependence on a state variable reflecting the history of sliding. The steady-state behavior is one of a positive (rate-strengthening) or negative (rate-weakening) dependence on the logarithm of the sliding rate. Such a state-variable formulation circumvents issues of ill-posedness of continuum models of faults whose strength is solely rate-dependent and a decreasing function of slip rate [e.g., \emph{Rice et al.}, 2001]. However, hypothetical models of stable fault slip allow for a wider range of potential descriptions, including a strictly rate-dependent formulation, which we explore in this work and further motivate in the following section. 



\subsection{Motivation and outline of current work}

We are interested in identifying key signatures of stable fault slip, including the spatiotemporal evolution of fault slip and slip rate in response to finite perturbations. At an elementary level, this involves the coupling of rate-strengthening fault shear strength with elastic deformation of the host rock. A rate-strengthening fault can be conceived to exist due to a number of linear or non-linear physical mechanisms, including rate- and state-dependent friction or a non-linear viscous response of a shear zone. Each non-linear formulation has the common feature of being linearizable about a finite or zero slip rate, tantamount to a linear viscous rheology. Here we consider this linear viscous response as the simplest rate-strengthening description. In doing so, we will find the universal asymptotic behavior for non-linear, rate-strengthening descriptions, the further analysis of which must be done on a case-by-case basis. Furthermore, in spite of being among the simplest mechanisms for stable, aseismic slip, the analysis of the mechanical response of faults with linearly viscous strength has received limited attention, apart from the works of \emph{Ida} [1974], \emph{Brener and Marchenko} [2002], and \emph{Ando et al.} [2012].

\

We take advantage of the linear form of the governing equations to analyze the transient response of a fault to space- and time-dependent loading. In Section 2 we introduce the equations governing problems in which slip occurs along an interface between elastic continua and in which the interfacial shear strength is proportional to slip rate. In Section 3 we introduce a non-dimensionalization and we draw comparisons with another problem: classical diffusion. We find that the two problems have direct analogies and we denote the problem governing in- or anti-plane viscous slip between elastic half-spaces as Hilbert diffusion; in Section 4 we provide the Green's function solution to this problem and use the Green's function approach to solve example initial value problems. In Section 5, we discuss the self-similar asymptotic expansion of such problems at large time and, in Section 6, we contrast the long-time behavior of large faults with that of finite-sized faults and spring-block models. We discuss our results in light of past observations and models in Section 7.

\begin{figure}[h]
\centering
 \noindent\includegraphics[scale=1.0]{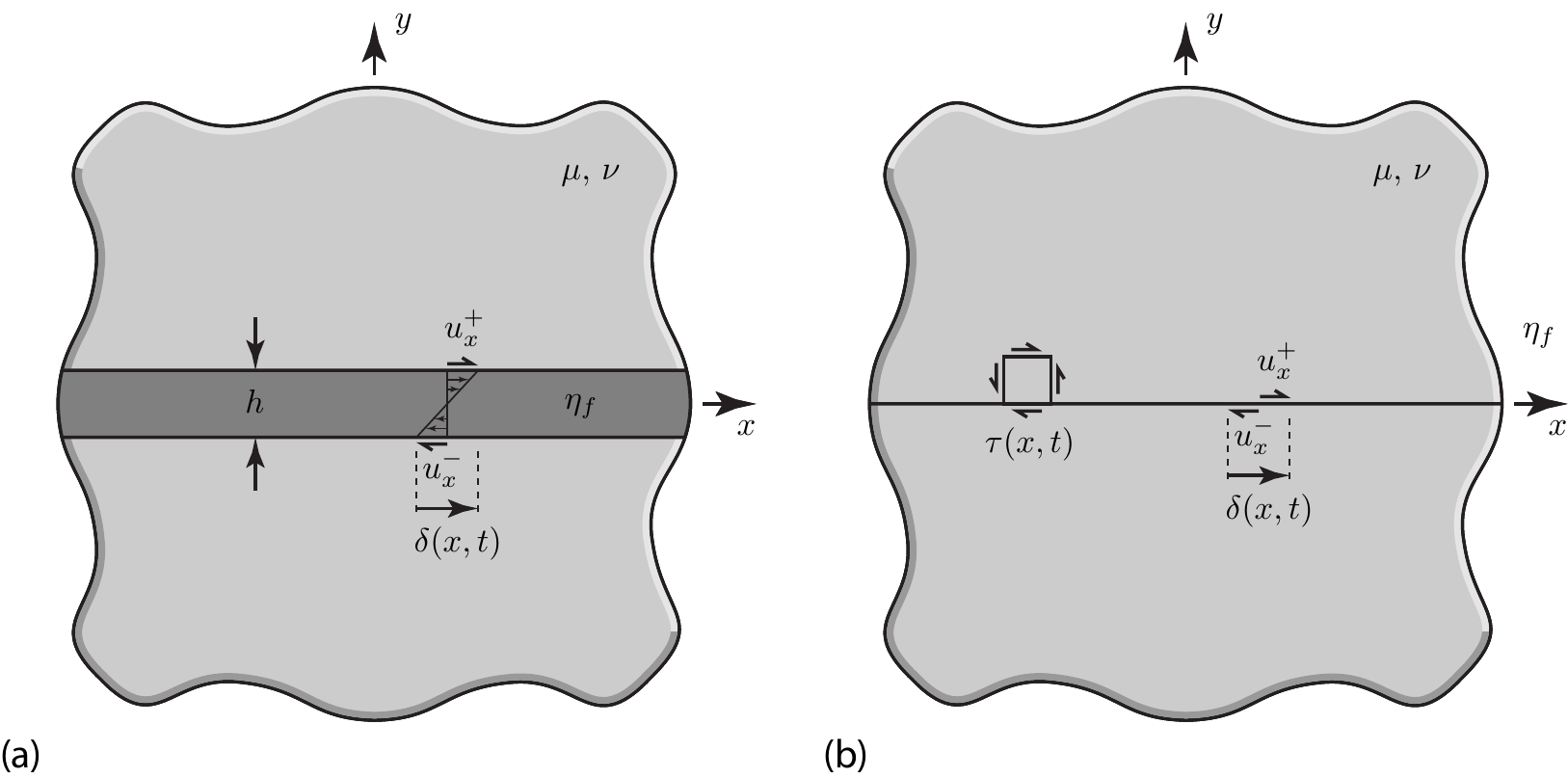}
 \caption{\small Illustration of model geometry and shear zone deformation. (a) A viscous shear zone of thickness $h$ and viscosity $\eta_f$ lies between two linear elastic half-spaces of shear modulus $\mu$ and Poisson ratio $\nu$. In-plane deformation of the shear zone is drawn, with the layer undergoing uniform strain and a relative displacement $\delta=u_x^+-u_x^-$, where $u_x^+$ and $u_x^-$ denote the displacement of the top and bottom faces of the shear zone. (b) Over distances much greater than $h$, the problem in (a) appears as two linear-elastic half-spaces in contact undergoing the relative displacement $\delta$ with that motion being resisted by an interfacial shear stress $\tau$.} \label{fig:setup}
 \end{figure}
 
\section{Governing equations}

We consider a fault shear zone to be a layer parallel to the $x$-$z$ plane with a uniform thickness $h$ in the fault-normal coordinate $y$ (Figure \ref{fig:setup}a). We presume that deformation is either within the $x$-$y$ plane (in-plane) or in the $z$-direction alone (anti-plane) and we denote $\tau$ as the stress component $\sigma_{yx}$ or $\sigma_{yz}$ within the shear zone for the in- and anti-plane cases, respectively. We presume inelastic deformation is localized to within the shear zone and that the variation of that deformation along $x$ occurs over a distance scale $d\gg h$. Consequently, scaling consideration of stress equilibrium within the shear zone implies that the shear stress $\tau$ satisfies $\partial\tau/\partial y=0$: $\tau$ is uniform within $|y|<h/2$ for a given $x$. We furthermore presume that the shear zone responds to deformation in linear viscous fashion, such that $\tau=\eta_f\dot\gamma$ where $\dot\gamma$ is twice the shear strain rate and $\eta_f$ is the shear zone's viscosity. Because $\tau$ is uniform in $y$, $\dot\gamma$ is uniform along $y$ as well. Owing to the condition that $d\gg h$, $\dot\gamma=\partial v_x/\partial y$ or $\partial v_z/\partial y$ for the in- and anti-plane strain cases, respectively, where $v_x$ and $v_z$ are the rates of displacement along the subscripted coordinates. Therefore, the displacement rate profile is as in Figure \ref{fig:setup}b and we may express the uniform strain rate in $y$ as $\dot\gamma(x,t)=V(x,t)/h$ where $V(x,t)$ is rate of relative displacement of the top end of the shear zone with respect to the bottom in the $x$ or $z$ direction: i.e., for in-plane deformation,  $V(x,t)=v_x(x,y=+h/2,t)-v_x(x,y=-h/2,t)$. The shear stress in the layer is then given by
\begin{equation}
\tau(x,t)=\eta_f \frac{V(x,t)}{h}
\label{eq:vi}
\end{equation}

We presume the material outside the layer responds in a linear elastic fashion, with shear modulus $\mu$ and Poission ratio $\nu$, to the internal inelastic deformation of the shear zone. The shear traction along the elastic bodies' boundaries at $y=\pm h/2$ is identically $\tau$, owed to the continuity of traction across the boundaries. We denote the relative displacement of the shear zone $\delta(x,t)$, such that  $V=\partial\delta/\partial t$, and for the in-plane case $\delta(x,t)=u_x(x,y=+h/2,t)-u_x(x,y=-h/2,t)$, where $u_x$ is the displacement in the $x$ direction. Given that we are concerned with variations of deformation of the shear zone along $x$ at scales $d\gg h$, we may effectively collapse the shear zone onto a fault plane along $y=0$ and no longer give the layer explicit consideration. Due to the quasi-static, elastic deformation of the material external to the shear zone, the distribution of $\tau$ along the fault surface (i.e., the surface formerly demarcated along $y=\pm h/2$, and now along $y=0^{\pm}$) must also satisfy
\begin{equation}
\tau(x,t)=\frac{\mu'}{\pi}\int_{-\infty}^{\infty}\frac{\partial\delta(s,t)/\partial s}{s-x}ds+\tau_b(x,t)
\label{eq:el}
\end{equation}
where $\mu'=\mu/[2(1-\nu)]$ and $\mu'=\mu/2$ for the in-plane and anti-plane cases, respectively. The last term on the right hand side is the shear tractions resolved on the fault plane due to external forcing while the first term is the change in shear tractions owed to a distribution of relative displacement, or slip, along the fault.

\

To draw useful comparisons later, we briefly consider here another configuration, one in which an elastic layer of thickness $b\gg h$ lies above the shear zone, and an elastic half-space lies underneath. In this case, when deformation along $x$ occurs over distances much longer than $b$, $\tau$ is instead given by [\emph{Viesca}, 2016, supplementary materials]
\begin{equation}
\tau(x,t)=E'b\frac{\partial^2\delta(x,t)}{\partial x^2}+\tau_b(x,t)
\end{equation}
where $E'=2\mu/(1-\nu)$ and $E'=\mu$ for the in-plane and anti-plane cases.

\

Combining (1) with (2) or (3) results in an equation governing the spatio-temporal evolution of slip $\delta$. In the section that follows we first non-dimensionalize this equation and subsequently highlight the diffusive nature of fault slip evolution.

\section{Problem non-dimensionalization}
Positing a characteristic slip rate $V_c$, we may in turn define characteristic values of shear stress $\tau_c=\eta_f V_c/h$, time $t_c=h/V_c$, along-fault distance $x_c=\mu' h/ \tau_c$ or $x_c=\sqrt{E' b h/ \tau_c}$, and slip $\delta_c=h$. We update our notation going forward to reflect the following nondimensionalization: 
$$V/V_c\Rightarrow V,\, \tau_b/\tau_c\Rightarrow \tau_b,\,  t/t_c\Rightarrow t,\, x/x_c\Rightarrow x,\, \text{ and }\delta/\delta_c\Rightarrow \delta$$

Doing so, the combination of (1) with (2) leads to 
\begin{equation}
\frac{\partial \delta}{\partial t}=\mathcal{H}\left(\frac{\partial\delta}{\partial x}\right)+\tau_b(x,t)
\label{eq:el}
\end{equation}
where we identify the operator $\mathcal{H}(f)=(1/\pi)\int_{-\infty}^\infty f(s)/(s-x)ds$ on a spatial function $f(x)$ as the Hilbert transform. Useful properties of the Hilbert transform are that $\mathcal{H}[\mathcal{H}(f)]=-f$ and that it commutes with derivatives in time and space: e.g., $d[\mathcal{H}(f)]/dx=\mathcal{H}[df/dx]$. 


\

For comparison, if we similarly combine and and non-dimensionalize (1) with (3), 
\begin{equation}
\frac{\partial \delta}{\partial t}=\frac{\partial}{\partial x}\left(\frac{\partial\delta}{\partial x}\right)+\tau_b(x,t)
\label{eq:eld}
\end{equation}
we immediately recognize that slip in this case satisfies the diffusion equation, with an external forcing term $\tau_b$. While (5) is a classical problem with known solution, the dynamics of (4), in contrast, has remained without comparable study despite being the simplest formulation of a rate-strengthening fault within an elastic continuum and despite also being among the simplest non-local, diffusion-type equations. However, in the sections to follow we highlight that the linearity of problem (4), which we refer to as the Hilbert diffusion equation, makes it as amenable to solution as the classical diffusion equation, though with several signature features.


\section{Solution via Green's function}
\label{sec:GFsoln}

We begin by looking for the fundamental solution, also known as the Green's function, to the problem in which the external forcing takes the form of an impulse  at position $x'$ and time $t'$, i.e., the function $G(x,t;x',t')$ satisfying
\begin{equation}
\frac{\partial G}{\partial t}=\mathcal{H}\left(\frac{\partial G}{\partial x}\right)+\delta_D(x-x')\delta_D(t-t')
\label{eq:GFeq}
\end{equation}
where we denote the Dirac delta function as $\delta_D(x)$. The impulse, represented by the last term in (\ref{eq:GFeq}), corresponds to a line load of unit magnitude momentarily applied to the fault.

\

Following standard techniques (outlined in Appendix A), the Green's function for the Hilbert diffusion equation is, for $t>t'$\begin{equation}
G(x,t;x',t')=\frac{1}{\pi(t-t')}\frac{1}{1+\left(\displaystyle\frac{x-x'}{t-t'}\right)^2}
\end{equation}
The Green's function exhibits self-similar behavior, in which distances $x$ stretch with time $t$. While not explicitly considered as such, the Green's function solution was effectively also derived by \emph{Ida} [1974] and \emph{Ando et al.} [2012]. To contrast, we recall that the Green's function solution for classical diffusion is
\begin{equation}
G(x,t;x',t')=\frac{1}{\sqrt{4 \pi(t-t')}}\exp\left[-\displaystyle \frac{(x-x')^2}{4(t-t')}\right]
\end{equation}
in which the power-law decay of the the Lorentzian $1/[\pi (1+s^2)]$ gives way to the exponential decay of a Gaussian and now distance $x$ stretches with $\sqrt{t}$. This latter change could be anticipated from scaling of the equations (4) and (5), excluding the source term, in which we find that [D]/[T] $\sim$ [D]/[L] in (4) and [D]/[T] $\sim$ [D]/[L$^2$] in (5), where the scalings of slip, time, or length are denoted by [D], [T], and [L], respectively.

\

Solutions to the problem of an arbitrary external forcing $\tau_b(x,t)$ are given by the convolution [e.g., \emph{Carrier and Pearson}, 1976]
\begin{equation}
\delta(x,t)=\int_{-\infty}^\infty\int_{-\infty}^t G(x,t;x',t')\tau_b(x',t')dt'dx'
\end{equation}
In the following section, we demonstrate the application of the Green's function solution to solve two elementary problems: a sudden and semi-infinite or localized step in stress on the fault.
 
\subsection{Example: Sudden step in stress}
\label{sec:fullsoln}
\begin{figure}[t]
\centering
 \noindent\includegraphics[scale=1.0]{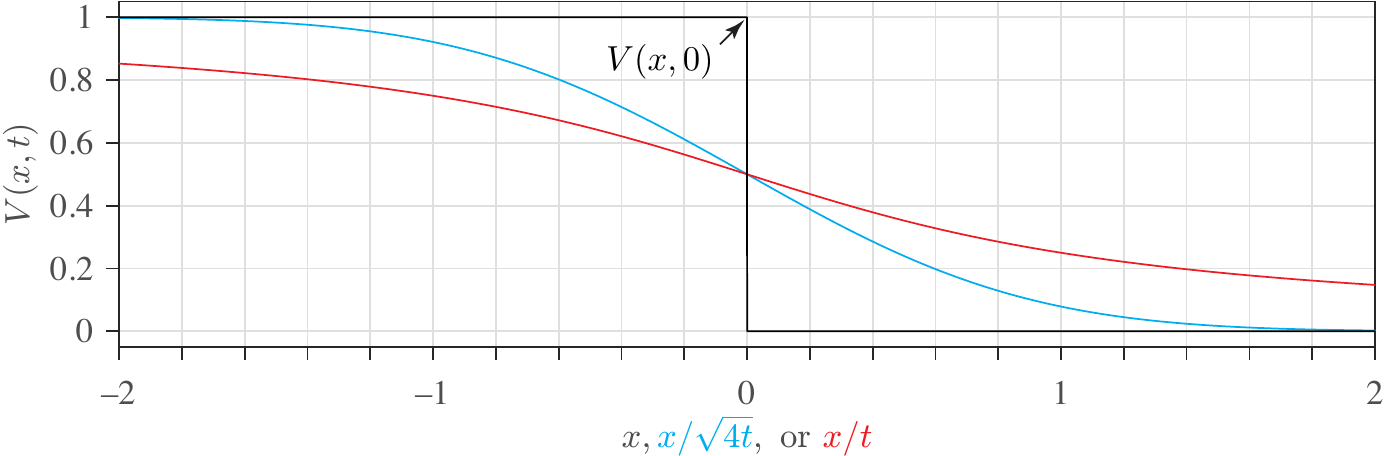}
 \caption{\small Similarity solutions (red and cyan) for slip rate $V$ for a linear viscous fault having undergone an initial step in stress on $x<0$, a problem equivalent to having the initial slip rate distribution shown in black. The cyan curve corresponds to the well-known error-function solution (\ref{eq:diffsimsoln}) satisfying the classical diffusion equation and exhibits characteristic exponential decay. In contrast, the red curve, illustrating the solution (\ref{eq:Hilbsimsoln}) satisfying the Hilbert diffusion equation, exhibits power-law decay with a distinct similarity variable, $x/t$.} \label{fig:simsoln}
 \end{figure}
 
As a simple example we consider the problem in which a sudden step in stress of unit magnitude is applied at $t=0$ and along $x<0$, or 
\begin{equation}
\tau_b(x,t)=H(-x)H(t)
\end{equation}
A solution is readily found by recognizing that $\partial^2\delta/(\partial x\partial t)$ also satisfies the Hilbert diffusion equation except the forcing now corresponds to $\partial^2\tau_b/(\partial x\partial t)=-\delta_D(x)\delta_{\textcolor{blue}{D}}(t)$, which is to within a sign of the Green's function problem with $x'=0$ and $t'=0$. Substituting for the slip rate $V=\partial \delta/\partial t$, the solution to this second problem is given by the Green's function, i.e. for $t>0$
\begin{equation}
\frac{\partial V}{\partial x} = -\frac{1}{\pi t} \frac{1}{1+(x/t)^2}
\end{equation}
and upon integration from $-\infty$ to $x$, with the condition that $V(-\infty,t>0)=1$ (i.e., the unit slip rate corresponding to the unit step in stress) we find that 
\begin{equation}
V(x,t)=\frac{1}{2}-\frac{1}{\pi}\arctan(x/t)=\frac{\text{arccot}(x/t)}{\pi}
\label{eq:Hilbsimsoln}
\end{equation}
Problem (10) can be equivalently posed as the initial value problem $V(x,0)=1$ for $x<0$ and $0$ otherwise. For comparison, the corresponding solution for the classical diffusion equation is 
\begin{equation}
V(x,t)=\frac{1}{2}-\frac{1}{2}\text{erf}(x/\sqrt{4t})=\frac{\text{erfc}(x/\sqrt{4 t})}{2}
\label{eq:diffsimsoln}
\end{equation}
We illustrate these similarity solutions, as well as their initial condition, in Figure \ref{fig:simsoln}.

\

A solution to a similar problem, in which $\tau_b(x,t)= H(-x)\delta_D(t)$, was also solved by \emph{Ida} [1974], using a complex variable approach. We can immediately see that, in the solution to this problem, slip takes the place of slip rate in (12), and that the problem can also be seen as the gradual smoothing of an initial dislocation of slip about $x=0$, suddenly placed at $t=0$. The solution procedure of \emph{Ida} [1974] identified  complex potential solutions to specific problems of (\ref{eq:el}), though appeared to overlook the self-similar nature of the solutions, the general Green's function solution procedure, and the diffusion connection.


\

\begin{figure}
\centering
 \noindent\includegraphics[scale=1.0]{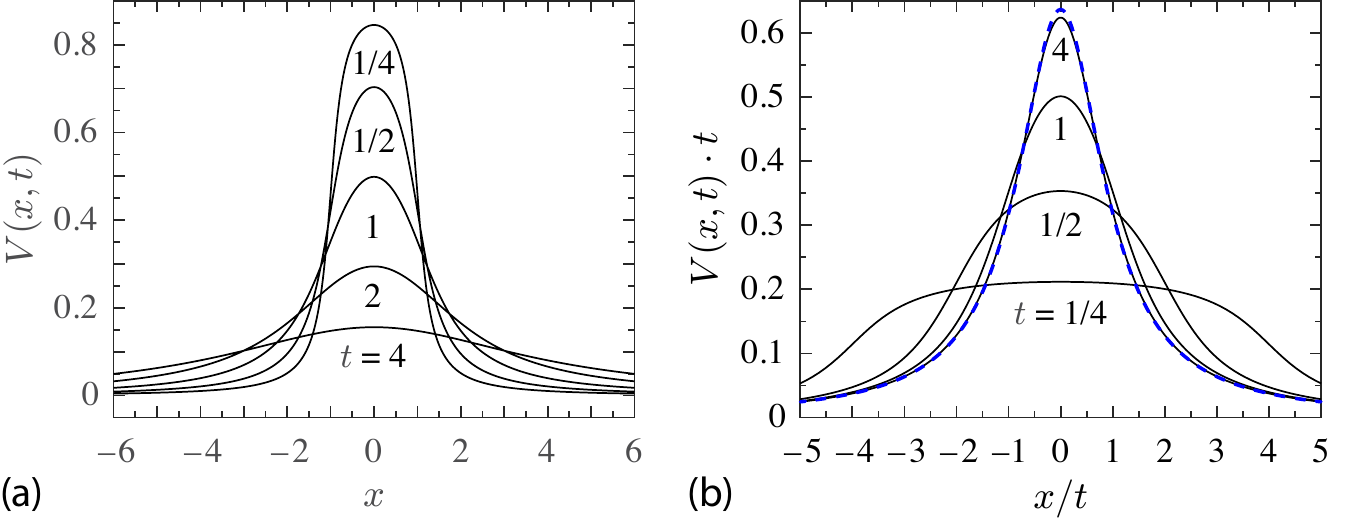}
 \caption{\small (a) Solutions for slip rate at intervals of time $t$ following an initial step in stress at $t=0$ with a boxcar spatial distribution about $x=0$. (b) Rescaling of the solutions in (a) to highlight the approach to the leading-order self-similar asymptotic solution (blue-dashed).} \label{fig:boxcar}
 \end{figure}

A simple extension of the problem is the localized step in stress, given by 
\begin{equation}
\tau_b(x,t)=B(x)H(t)
\end{equation}
where the boxcar function $B(x)=1$ on $|x|<1$ and 0 otherwise. The corresponding solution can be found using superposition of the pair of solutions to the problem (10), shifted to $x=\pm 1$
\begin{equation}
V(x,t)=\frac{\arctan\left[(x+1)/t\right]-\arctan\left[(x-1)/t\right]}{\pi}
\label{eq:boxcar}
\end{equation}
Figure \ref{fig:boxcar}a shows the solution at intervals in time.

\section{Self-similar asymptotics}

Solutions to initial value problems exhibit self-similar behavior at long time: i.e., an asymptotic expansion of solutions may be written in the limit $t\rightarrow \infty$ (Appendix B),
\begin{equation}
V(x,t)=\sum_{n=1}^\infty \frac{1}{t^n} \text{Re}[c_n f_n(\eta)] \quad \text{with }\eta=x/t
\label{eq:exp}
\end{equation}
 where the complex constants $c_n=a_n-i b_n$ and $f_n$ are the complex functions
\begin{equation}
f_n(\eta)=\frac{i}{\pi(\eta+i)^n}
\end{equation}
the real and imaginary parts of which are listed in Table 1 for the first several $n$. For initial conditions that vanish sufficiently fast outside of a finite region, the constants $a_n$ and $b_n$ are determined by the ($n-1$)th moment of the initial distributions of $V$ and $\mathcal{H}(V)$ (Appendix C)

\begin{align}
a_n&= \int_{-\infty}^{\infty}x^{n-1}V(x,0)dx \label{eq:an}\\
b_n&=\int_{-\infty}^\infty x^{n-1}\mathcal{H}[V(x,0)]dx \label{eq:bn}
\end{align}

For sufficiently localized steps in stress, such as the first example in the following sub-section, the asymptotic expansion reduces to the series
\begin{equation}
V(x,t)=a_1\frac{g_1(\eta)}{t}+a_2\frac{g_2(\eta)}{t^2}+O(t^{-3})
\end{equation}
where $g_n(\eta)$ are the real parts of $f_n(\eta)$. However, due to the long-range nature of elastic interactions in the crust, physical processes of interest will not typically have such a spatially compact loading. In the sub-sections to follow we will also consider two examples in which the external forcing exhibits a slow decay in space. The first such example is a non-local step in stress whose solution follows simply from the localized loading problem considered. The last example examines an increase in fault slip rate induced by a specific physical process: the sudden slip of a small, secondary fault near a larger, rate-strengthening fault. This last example highlights additional considerations necessary to arrive to the appropriate asymptotic expansion.

\subsection{Example: Asymptotic behavior of a sudden step in stress}
\label{sec:boxcarasy}
Here we look for the asymptotic expansion for the problem (14), which corresponds to the initial value problem $$V(x,0)=B(x)$$ where $B(x)$ is the boxcar function introduced previously. For this case, the $(n-1)$th moments of the initial distribution are 
\begin{equation}
a_n=\int_{-1}^1x^{n-1}dx=\begin{cases}
 2/n& n\text{ odd}\\
0 & n\text{ even}
\end{cases}
\end{equation}
To calculate the $(n-1)$th moments of $\mathcal{H}[V(x,0)]$, we first find $\mathcal{H}[B(x)]$. To do so, we use the commutative property of the Hilbert transform
\begin{equation}
\frac{d}{dx}\left(\mathcal{H}[B(x)]\right)=\mathcal{H}[B'(x)]=\mathcal{H}[\delta_D(x+1)-\delta_D(x-1)]=\frac{1}{\pi}\frac{1}{-1-x}-\frac{1}{\pi} \frac{1}{1-x}
\end{equation}
and subsequently integrate to find that
\begin{equation}
\mathcal{H}[B(x)]=\frac{1}{\pi}\ln\left |\frac{1-x}{1+x}\right |
\end{equation}
It then follows that the moments $b_n$ do not exist as the integrals are divergent. Setting $b_n=0$, the asymptotic expansion is
\begin{equation}
V(x,t) = \frac{2}{\pi t}\frac{1}{1+(x/t)^2} + \frac{2/3}{\pi t^3}\frac{3(x/t)^2-1}{\left[1+(x/t)^2\right]^3}+ \frac{2/5}{\pi t^5}\frac{5(x/t)^4-10(x/t)^2+1}{\left[1+(x/t)^2\right]^5}+O(t^{-7})
\end{equation}
In Figure \ref{fig:boxcar}b we show that the general problem solution given by (\ref{eq:boxcar}) converges to the leading-order asymptotic term.

\begin {table}[h]
\caption{\small Real and imaginary parts of $f_n(\eta)$ for $n=1,...,4$} \label{tab1} 

\

\centering

\small\begin{tabular}[c]{ c  c  c }
$n$& $\text{Re}[\pi f_n(\eta)]$ & $\text{Im}[\pi f_n(\eta)]$ \\
\hline 
 $1$ & $\displaystyle \frac{1}{1+\eta^2}$ & $\displaystyle \frac{\eta}{1+\eta^2}$ \\
\hline
 $2$ & $\displaystyle \frac{2\eta}{(1+\eta^2)^2}$ & $\displaystyle \frac{\eta^2-1}{(1+\eta^2)^2}$ \\
\hline
$3$ & $\displaystyle \frac{3\eta^2-1}{(1+\eta^2)^3}$ & $\displaystyle \frac{\eta^3-3\eta}{(1+\eta^2)^3}$ \\
\hline
$4$ & $\displaystyle\frac{4(\eta^3-\eta)}{(1+\eta^2)^4}$ & $\displaystyle \frac{\eta^4-6\eta^2+1}{(1+\eta^2)^4}$ \\
\hline
\end{tabular}
\end{table}

\subsection{Example: A complementary problem}
To contrast with the preceding example, determining the asymptotic behavior of an initial value problem such as
\begin{equation}
V(x,0)=-\frac{1}{\pi}\ln\left |\frac{1-x}{1+x}\right |
\end{equation}
may at first appear to be problematic as the moments $a_n$ do not exist. However, we recall a result from the preceding example to note that
\begin{equation}
\mathcal{H}[V(x,0)]=B(x)
\end{equation}
Additionally, since $\mathcal{H}[V(x,t)]$ also satisfies the Hilbert diffusion equation (\ref{eq:el}), the asymptotic behavior of $\mathcal{H}(V)$ is then precisely that of $V$ in the preceding example
\begin{equation}
\mathcal{H}[V(x,t)] = \frac{2}{\pi t}\frac{1}{1+(x/t)^2} + \frac{2/3}{\pi t^3}\frac{3(x/t)^2-1}{\left[1+(x/t)^2\right]^3} + \frac{2/5}{\pi t^5}\frac{5(x/t)^4-10(x/t)^2+1}{\left[1+(x/t)^2\right]^5}+O(t^{-7})
\end{equation}
Upon taking the Hilbert transform of both sides, we find the asymptotic decay of slip rate to follow
\begin{equation}
V(x,t) = \frac{2}{\pi t}\frac{x/t}{1+(x/t)^2} + \frac{2/3}{\pi t^3}\frac{(x/t)^3-3(x/t)}{\left[1+(x/t)^2\right]^3} + \frac{2/5}{\pi t^5}\frac{(x/t)^5-10(x/t)^3+5(x/t)}{\left[1+(x/t)^2\right]^5}+O(t^{-7})
\label{eq:comp}
\end{equation}

\

Alternatively, we may have directly calculated coefficients $b_n$ (\ref{eq:bn})
\begin{equation}
b_n=\int_{-1}^1x^{n-1}dx=\begin{cases}
 2/n& n\text{ odd}\\
0 & n\text{ even}
\end{cases}
\end{equation}
and with $a_n=0$ retrieved the same expression (\ref{eq:comp}) from (\ref{eq:exp}).

\begin{figure}
\centering
 \noindent\includegraphics[scale=1.0]{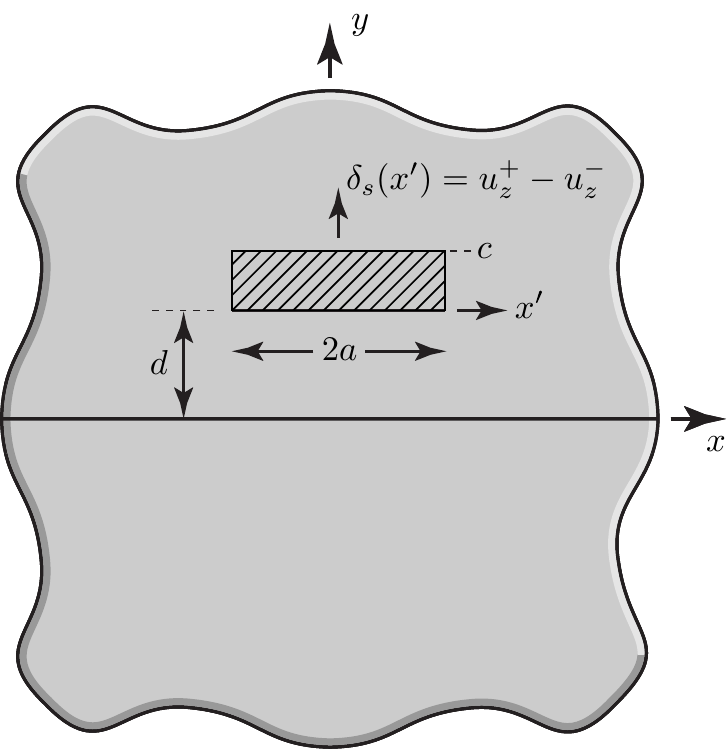}
 \caption{\small Illustration of the model problem considered in Section \ref{sec:dipole}, a relatively small secondary fault undergoes an amount $c$ of relative anti-plane displacement $\delta_s$ at $t=0$. The secondary fault has a width $2a$ and is located a distance $d$ away from the principal fault, which lies along the $x-z$ plane.} \label{fig:secondfault}
 \end{figure}
\subsection{Example: Dislocation dipole near a principal fault}
\label{sec:dipole}

We now consider the problem of sudden anti-plane slip within a region near a principal fault, as in Figure \ref{fig:secondfault}. Specifically, we imagine a secondary fault of length $2a$ lies parallel to the principal fault and a distance $d$ away. At $t=0$ the secondary fault suddenly slips an amount $c$, which is represented by the sudden appearance of a pair of screw dislocations with Burgers vectors of opposite signs a distance $2a$ apart, comprising a dislocation dipole. We are interested in determining the response of the principal fault to this perturbation. The off-fault dislocation dipole induce the stress change on the principal fault of the form
\begin{equation}
\tau_b(x,t)=\frac{c}{\pi}\left[\frac{x-a}{d^2+(x-a)^2}-\frac{x+a}{d^2+(x+a)^2}\right] H(t)
\end{equation}
We may alternatively consider the posed problem as an initial value problem for the anti-plane slip rate of the principal fault
\begin{equation}
V(x,0)=\frac{c}{\pi}\left[\frac{x-a}{d^2+(x-a)^2}-\frac{x+a}{d^2+(x+a)^2}\right]
\end{equation}

\

One path to the full solution is to recognize that the function $h(x,t)=\mathcal{H}[V(x,t)]$ also satisfies the Hilbert diffusion equation. We may now pose the problem as an inital value problem for $h(x,t)$ where
\begin{equation}
h(x,0)=\mathcal{H}[V(x,0)]=\frac{c}{\pi}\left[\frac{d}{d^2+(x-a)^2}-\frac{d}{d^2+(x+a)^2}\right]
\end{equation}
Recognizing the form of the Green's function, the solution of the auxiliary problem is, by inspection,
\begin{equation}
h(x,t)=\frac{c}{\pi(t+d)}\left[\frac{1}{1+\left[(x-a)/(t+d)\right]^2}-\frac{1}{1+\left[(x+a)/(t+d)\right]^2}\right]
\end{equation}
The solution to the original problem then follows from the inversion for $V(x,t)=-\mathcal{H}[h(x,t)]$
\begin{equation}
V(x,t)=\frac{c}{\pi(t+d)}\left[\frac{(x-a)/(t+d)}{1+[(x-a)/(t+d)]^2}-\frac{(x+a)/(t+d)}{1+[(x+a)/(t+d)]^2}\right]
\end{equation}
where in the above we used the known transform $\mathcal{H}[x/(b^2+x^2)]=b/(b^2+x^2)$ for $b$  a constant in space and the property that $\mathcal{H}\{\mathcal{H}[f(x)]\}=-f(x)$.

\

To find the asymptotic behavior, we look to determine the coefficients of the expansion, $a_n$ and $b_n$. However, this example is unlike the preceding ones owing to the power-law decay of both the initial slip rate and its Hilbert transform. Consequently, the moment integral expressions for $a_n$ and $b_n$ as written in (\ref{eq:an}) and (\ref{eq:bn}) are divergent for $n$ sufficiently large. Nonetheless, we may proceed to determine the coefficients as follows. 

\

First, given the problem symmetry, $V(x,t)=V(-x,t)$, we may anticipate that $a_n=0$ when $n$ is even and $b_n=0$ when $n$ is odd as we expect to keep only functions that are symmetric about $x=0$ in the asymptotic expansion. Proceeding to determine the moments of the initial distribution and its Hilbert transform, leads to $a_1 = 0$ and $b_2=2ac$, the last of which we identify as the net moment of the dislocation dipole

\begin{equation*}
m_o=2ac
\end{equation*}
Examining the next-order moment $a_3$, we find that the integral is divergent. This divergence is owed to the integrand approaching a finite value as $x\rightarrow\pm\infty$:
\begin{equation*}
\lim_{x\rightarrow\pm\infty} x^2V(x,0) = \frac{m_o}{\pi}
\end{equation*}
Subtracting this limit value from the integrand when calculating the moment, we find
\begin{equation*}
a_3=\int_{-\infty}^\infty \left[x^2V(x,0) -\frac{m_o}{\pi}\right]dx=-m_o(2d)
\end{equation*}
The integrand of the moment $b_4$ likewise diverges due to the integrand having a non-zero limit in the far-field 
\begin{equation*}
\lim_{x\rightarrow\pm\infty} x^3\mathcal{H}[V(x,0)] = -\frac{m_o(2d)}{\pi}
\end{equation*}
Proceeding similarly, we find that
\begin{equation*}
b_4=\int_{-\infty}^\infty \left[x^3\mathcal{H}[V(x,0)] +\frac{m_o (2d)}{\pi}\right]dx=m_o(a^2-3d^2)
\end{equation*}
To determine the next-order coefficient $a_5$, we note that as $x\rightarrow\infty$
\begin{equation*}
x^4V(x,0)\approx x^2\frac{m_o}{\pi}+\frac{m_o(a^2-3d^2)}{\pi} + O(x^{-2})
\end{equation*}
the first two terms of which preclude the moment integral from being convergent. Subtracting these terms from the moment integrand, we arrive to
\begin{equation*}
a_5=\int_{-\infty}^\infty \left[x^4V(x,0) -x^2 \frac{m_o}{\pi}-\frac{m_o(a^2-3d^2)}{\pi}\right]dx=m_o[4d(d^2-a^2)]
\end{equation*}

\

Thus, the four leading-order terms in the asymptotic expansion are
\begin{align}
V(x,t)=\,&\frac{m_o}{\pi t^2}\frac{(x/t)^2-1}{[1+(x/t)^2]^2}-\frac{m_o (2 d)}{\pi t^3}\frac{3(x/t)^2-1}{[1+(x/t)^2]^3} + \frac{m_o(a^2-3d^2)}{\pi t^4}\frac{(x/t)^4-6(x/t)^2+1}{[1+(x/t)^2]^4}\notag\\[9 pt]
+\,&\frac{m_o[4d(d^2-a^2)]}{\pi t^5}\frac{5(x/t)^4-10(x/t)^2+1}{[1+(x/t)^2]^5}+ O(t^{-7})
\end{align}
We emphasize that the leading term is independent of the distance $d$ and dependent on $a$ and $c$ only insofar as they determine the dipole moment $m_o$. Subsequent terms are found following the procedure outlined above, i.e., given that $a_1=0$ and $b_2=m_o$
\begin{align}
a_n &=\int_{-\infty}^{\infty} \left[ x^{n-1}V(x,0) -\frac{1}{\pi}\sum_{k=1}^{(n-1)/2}x^{(n-1)-2k}\,b_{2k} \right]dx\quad \text{for $n=3,5,7,...$ }\\[9 pt]
b_n &= \int_{-\infty}^{\infty} \left[ x^{n-1}\mathcal{H}[V(x,0)] -\frac{1}{\pi}\sum_{k=1}^{(n-2)/2}x^{(n-2)-2k}\,a_{2k+1} \right]dx \quad \text{for $n=4,6,8,...$ }
\end{align}

\section{Comparison with spring-block and finite-fault models}
\label{sec:sprfinite}

In the preceding section we found that the asymptotic response to a stress step is a power-law decay in slip rate with time. In this section we highlight that such power-law decay transitions to exponential decay in time if locked fault boundaries are encountered. This transition to a more rapidly decaying slip rate is owed to limitations of the compliance of the coupled fault-host rock system. We begin by considering a simple system with a limited compliance: a sliding block attached to a spring with fixed stiffness, finding the exponential decay of slip rate following a step in stress. We subsequently examine a continuum system with limited compliance: a finite fault embedded within a full space. We examine the response of such a fault to a step in stress and show that the decay of slip rate is expressible as the sum of orthogonal modes whose amplitudes have an exponential decay with time. We also highlight that the conditions for such asymptotic behavior are rather strict, requiring slip not penetrate beyond a particular position, and that non-exponential decay can be expected if fault boundaries are modeled as transitions in fault rheology.

\subsection{Spring-block model}
\label{sec:sprblock}

We consider a rigid block sliding on a rigid substrate. The block is attached to one end of a spring, with stiffness k, the other end of which is attached to a oad point pulled at a constant rate $V_p$. The basal shear stress is denoted $\tau$ and $\tau_b$ is the shear stress applied to the top of the block. For quasi-static deformation, 
\begin{equation}
\tau(t)=k[V_p t-\delta(t)]+\tau_b(t)
\end{equation}
where $\delta(t)$ is both the basal slip and the displacement of the block. As before, we presume the basal interface is modeled as a thin viscous layer such that the interfacial shear strength is
\begin{equation}
\tau_s(t)=\eta\frac{V(t)}{h}
\label{eq:SBfric}
\end{equation}

We consider the response to a sudden step in applied stress at $t=0$
\begin{equation}
\tau_b(t)=\Delta\tau H(t)
\end{equation}
With the initial condition $\delta(0)=0$, the block's displacement relative to the load point is given by
\begin{equation}
\delta(t)-V_p t=(A-B)[1-\exp(-t/t_c)]
\end{equation}
where $t_c=\eta/(hk)$, $A=\Delta\tau/ k$, and $B=V_p t_c$


\subsection{Finite-fault model}
\label{sec:FF}
We now consider the in-plane or anti-plane slip of a fault with a finite length $2L$. The dimensional shear stress on the fault plane is given by. 
\begin{equation}
\tau(x,t)=\frac{\mu'}{\pi}\int_{-L}^L\frac{\partial \delta(s,t)/\partial s}{s-x}ds+\tau_b(x,t)
\end{equation}
with the condition that there is no slip for $|x|>L$, including the endpoints: $\delta(\pm L,t)=0$.

\

Nondimensionalizing as done previously to arrive to (\ref{eq:el}), 
\begin{equation}
\tau(x,t)=\frac{1}{\pi}\int_{-\bar L}^{\bar L}\frac{\partial \delta(s,t)/\partial s}{s-x}ds+\tau_b(x,t)
\label{eq:FF}
\end{equation}
where the dimensionless fault length is denoted $\bar L=L/x_c$. Requiring that $\tau=\tau_s$ within the fault plane $|x|<\bar L$, the time rate of the above is 
\begin{equation}
\frac{\partial V(x,t)}{\partial t}=\frac{1}{\pi}\int_{-\bar L}^{\bar L}\frac{\partial V(s,t)/\partial s}{s-x}ds+\frac{\partial \tau_b(x,t)}{\partial t}
\label{eq:FFrate}
\end{equation}
with the condition that $V(\pm \bar L,t)=0$.

\

We now look to determine how the finite fault responds to a step in stress $\tau_b(x,t)=f(x)H(t)$. We may alternatively pose this problem as the initial value problem for the slip rate as $V(x,0)= f(x)$ where the slip rate for $t>0$ satisfies
\begin{equation}
\frac{\partial V(x,t)}{\partial t}=\frac{1}{\pi}\int_{-\bar L}^{\bar L}\frac{\partial V(s,t)/\partial s}{s-x}ds
\label{eq:ff}
\end{equation}
We begin by looking for solutions to (\ref{eq:ff}) of the form
\begin{equation}
V(x,t)=\omega(x/\bar L)\exp(\lambda t/\bar L)
\end{equation}
which, when substituted into (\ref{eq:ff}) leads to the eigenvalue problem 
\begin{equation}
\lambda \omega(x)=\frac{1}{\pi}\int_{-1}^1\frac{ d\omega(s)/ds}{s-x}ds
\end{equation}
previously analyzed in the context of earthquake nucleation on linearly slip-weakening faults [\emph{Dascalu et al., 2000}; \emph{Uenishi and Rice}, 2003]. As discussed by those authors, solutions to the eigen equation have the following properties: the eigenmodes $\omega_n$ are orthogonal and correspond to a set of discrete, unique eigenvalues $\lambda_n<0$, which can be arranged in decreasing order $0>\lambda_1>\lambda_2 ...$ . Consequently, the initial velocity distribution can be decomposed into a sum of the eigenfunctions 
\begin{equation}
V(x,0)=\sum_{n=1}^\infty v_n\omega_n(x/\bar L)
\end{equation}
where, owing to the orthogonality of $\omega_n$, the coefficients of the expansion are given by
\begin{equation}
v_n=\displaystyle\frac{\displaystyle\int_{-1}^1V(s\bar L,0)\omega_n(s)ds}{\displaystyle\int_{-1}^1 \omega^2_n(s)ds}
\label{eq:coeff}
\end{equation}
and consequently, the solution to the initial value problem is
\begin{equation}
V(x,t)=\sum_{n=1}^\infty v_n \omega_n(x/\bar L)\exp(\lambda_n t/\bar L)
\label{eq:FFsoln}
\end{equation}
The most slowly decaying symmetric and anti-symmetric modes can be accurately approximated, to within 1\% besides an arbitrary scale factor, as
\begin{align} \omega_1(x)&\approx\sqrt{1-x^2}(1-x^2/3) \label{eq:omeg1}\\
\omega_2(x)&\approx\sqrt{1-x^2}[1-(4x/5)^2]x
\end{align}
and have the eigenvalues $\lambda_1=-1.157774...$ and $\lambda_2=-2.754755...$ . The eigenfunctions and eigenvalues may be solved for numerically, as done by \emph{Dascalu et al} [2000] using a Chebyshev polynomial expansion, \emph{Uenishi and Rice} [2000] using a discrete dislocation technique, or \emph{Brantut and Viesca} [2015] using Gauss-Chebyshev-type quadrature for a comparable problem. Doing so, (\ref{eq:coeff}) and (\ref{eq:FFsoln}) are then readily evaluated numerically. The eigenfunction expansion can also be used to construct a Green's function solution to the problem (\ref{eq:FF}) or (\ref{eq:FFrate}) with $\tau_b$ or $\partial \tau_b/\partial t$ being equal to $\delta(x-x')\delta(t-t')$, which we defer to later work.

\

The finite-fault results here show what may be the fate of the self-similar spreading of slip rate discussed in the preceding sections, which presumed that rigid boundaries are not yet encountered. Specifically, the self-similar behavior may be seen as intermediate: relevant when distributions of elevated slip rate extend over distances comparable to or greater than the initiating source, but still lie some distance away from locked boundaries. If such boundaries are encountered, the self-similar behavior transitions to one in which slip rate no longer spreads and instead decays in place exponentially with time, akin to the behavior seen in spring-block models. In the following section we will show that self-similar spreading implies a constant moment rate following a stress step. We can immediately deduce from (\ref{eq:FFsoln}) that the moment rate, an integral of slip rate over the fault, will ultimately exponentially decay back to zero from this constant value once locked boundaries begin to influence slip rate within the fault interior. The rigidly locked boundary condition may accurately represent the termination of a fault segment, in which a fault discontinuity transitions to intact rock. However, a more accurate boundary condition for mature faults, which may slow down the abrupt, exponential decay of slip rate, may be a rheological transition accompanying increases in temperature down dip, or heterogeneity in fault material properties along strike.

\section{Discussion}

\subsection{Relation with existing spring-block and continuum models for postseismic slip}

Observations of surface displacements that appear to increase logarithmically with time have been used as evidence for postseismic deformation due to frictional afterslip or ductile deformation of faults, in part owed to the observation that such a logarithmic time dependence arises in models of stable fault slip. Early models have focused on the displacement response of spring-block systems to steps in stress, and a logarithmic increase of displacement has been shown to arise in models that include an explicit dependence of strength on the logarithm of the sliding rate, including slip rate- and state-dependent strength descriptions [e.g., \emph{Marone et al.} 1991; \emph{Perfettini and Avouac}, 2004; \emph{Montesi}, 2004; \emph{Helmstetter and Shaw}, 2009]. In general, however, spring-block models in which the frictional strength linearizes about a finite or zero value of slip rate---including slip rate- and state-dependent friction---all exhibit a slip rate that, at sufficiently long time, relaxes back to a zero or finite value exponentially with time following a step in stress, under a negligible or finite value of the load-point velocity $V_p$. Nonetheless, a logarithmic growth of displacement can arise from a spring-block model in some limiting cases when considering a frictional formulation that does not linearize about $V=0$, specifically a strength relation $\tau_s(V)=\tau_o+c \ln(V)$. Revisiting the analysis of Section \ref{sec:sprblock} for the response of a spring-block system to a sudden step in stress and replacing the linear viscous strength relation with this logarithmic slip-rate dependence, one finds that the asymptotic behavior is $V\sim1/t$, and hence $\delta\sim\ln(t)$, provided that the load point velocity $V_p=0$. For finite values of $V_p$, the slip rate approaches a $1/t$ decay only as an intermediate asymptote at sufficiently short times after the step, followed by a transition to an exponential decay of slip rate to the finite value of $V_p$ at long time.

\

That strength may depend on the logarithm of slip rate is consistent with experimental observations and the constitutive formulation for rate- and state-dependent friction insofar as these both exhibit a logarithmic dependence under steady-sliding conditions [\emph{Dieterich}, 1978, 1979; \emph{Ruina}, 1980, 1983]. Considering \textcolor{red}{that} the full rate- and state-dependence, with aging-law state evolution, does not qualitatively change the behavior following a step in stress applied to a spring-block system in comparison with a strictly logarithmic dependence on slip rate: a $1/t$ decay shortly after the step transitions to oscillatory exponential decay at long time [e.g., \emph{Rice and Ruina}, 1983; \emph{Helmstetter and Shaw}, 2009]. This has made the spring-block model a common representation of post-seismic fault slip when attempting to fit observed displacements with the functional form $\delta(t)=\delta_c\ln(1+t/t_c)$, and the goodness of fit has furthermore been taken as evidence in support of a rate- and state-dependent description governing fault strength in the post-seismic period. Specifically, the spring-block model of \emph{Marone et al.} [1991] is based on the strict logarithmic dependence and $V_p=0$, while \emph{Perfettini and Avouac} [2004] relax the latter assumption and \emph{Helmstetter and Shaw} [2009] consider the full rate- and state-dependent formulation and show the limiting cases where a $1/t$ decay in slip rate may arise.

\

However, we find that a logarithmic dependence on time of both displacement on the fault---as well as that at a free surface---also emerges from a fault with a simple linear viscous strength. This emergence is owed to the interactions between points on the fault that accompany a continuum description, which is neglected in spring block models. The logarithmic dependence is readily seen for fault displacement given the asymptotic expansion for slip rate (\ref{eq:exp}) following a step in stress on the fault, which at the leading order has $a_1\neq0$ and $b_1=0$, and provides the asymptotic form of fault slip upon integration
\begin{equation}
\delta(x,t)=\delta_0+\frac{a_1}{\pi}\ln\left(x^2+t^2\right)+O\left(t^{-1}\right)
\end{equation}
where $\delta_0$ is a constant.

\


\begin{figure}[t]
\centering
 \noindent\includegraphics[scale=1.0]{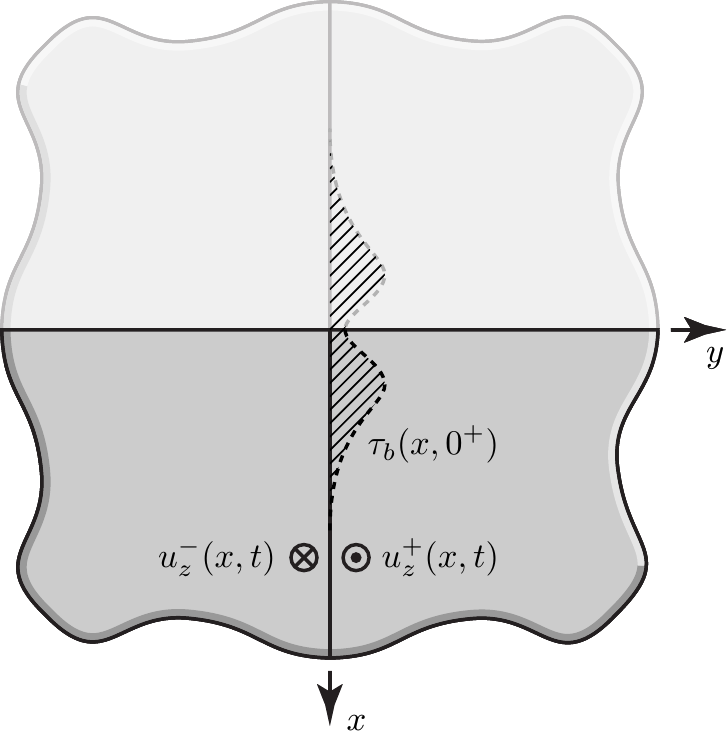}
 \caption{\small An anti-plane fault in the $x$-$z$ plane intersects a free surface at $x=0$. The solution to a sudden step in stress $\tau_b$ at $t=0^+$ along the fault ($x>0$) is found by method of images. The image problem is shown as a transparent continuation in the upper half plane, $x<0$. The hatched area represents the net force exerted on the imaged fault per unit distance along strike, which determines the leading-order coefficient of the asymptotic expansion, $a_1$.} \label{fig:halfspace}
 \end{figure}
 
\

How will the asymptotic decay of slip rate on the fault be reflected at displacement on the surface? We consider a simple model of a long, vertical fault that undergoes strike-slip (anti-plane) motion and intersects the free surface, lying along $x\geq 0$ (Figure \ref{fig:halfspace}). For simplicity, we imagine that the fault has undergone a sudden step in shear stress with some distribution along $x\geq 0$. The solution to this problem is found by method of images [e.g., \emph{Segall}, 2010], in which the original problem is reflected about $x=0$ such that the fault now lies in a full space with a step in stress symmetric about $x=0$. The solution for the half-space problem is the full-space solution for $x\geq 0$. Given the fault slip rate $V(x,t)$, the dimensionless out-of-plane displacement rate off the fault is
\begin{equation}
v_z(x,y,t)=\frac{1}{2\pi}\int_{-\infty}^\infty\frac{\partial V(s,t)}{\partial s}\arctan\left(\frac{s-x}{y}\right)ds
\label{eq:surfrate}
\end{equation}
which follows from the superposition of the solution for the displacement field surrounding a screw dislocation in an infinite medium [e.g., \emph{Hirth and Lothe}, 1982]. Given that the fault slip rate for this problem decays at long time as
\begin{equation}
V(x,t)=\frac{a_1}{\pi t}\frac{1}{1+(x/t)^2}+\frac{a_3}{\pi t^3}\frac{3(x/t)^2-1}{[1+(x/t)^2]^{3}}+O(t^{-5})
\label{eq:Vstepasym}
\end{equation}
the substitution of the spatial derivative of this expansion into (\ref{eq:surfrate}) and the evaluation of the integral for positions at the surface ($x=0$) reveals that the displacement rate there decays as
\begin{equation}
v_z(0,y,t)=\frac{a_1}{2\pi}\frac{1}{(y+t)}+O(t^{-3})\label{eq:disprate}
\end{equation}
and the long-time behavior of the surface displacement follows from the time integration of (\ref{eq:disprate})
\begin{equation}
u_z(0,y,t)=u_0+\frac{a_1}{2\pi}\ln(y+t)+O(t^{-2})
\label{eq:surfdispasy}
\end{equation}
where $u_0$ is a constant of integration. Here we see that the expected long-time evolution of the surface displacement is logarithmic with time. As we will come to see in the following section, the coefficient of the logarithm $a_1$ represents the net force exerted (per unit distance perpendicular to strike) on the fault by the sudden increase in shear stress.

\

\begin{figure}[t]
\centering
 \noindent\includegraphics[scale=1.0]{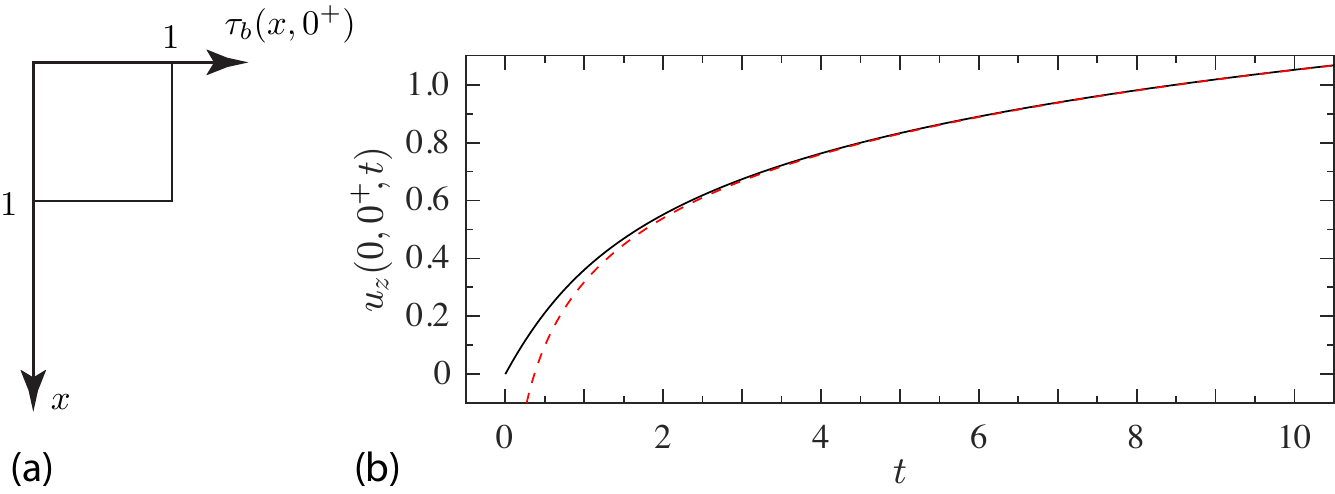}
 \caption{\small (a) Illustration of a boxcar step in stress with depth $x$ on an anti-plane (strike-slip) fault, beginning at $t=0^+$. (b) Evolution of the anti-plane displacement component $u_z$ at the free surface ($x=0$) to one side of the fault ($y=0^+$), given by (\ref{eq:exsurfdisp}) (black), following the step in fault shear stress; the displacement quickly approaches the logarithmic asymptote (\ref{eq:exsurfdispasy}) (red-dashed).} \label{fig:exhalfspace}
 \end{figure}

We highlight how quickly the logarithmic time dependence is approached by using a simple initial stress step example: $\tau_b(x,t)=B(x)H(t)$ for $x>0$. The full slip rate solution for the half-space follows from the example problem considered previously in Section \ref{sec:GFsoln} for the full space and is given by (\ref{eq:boxcar}) with $x>0$. Using this, we directly calculate the displacement history on one side of the fault $y=0^+$ at the surface $x=0$
\begin{align}
u_z(0,0^+,t)&=\int_0^t v_z(0,0^+,t)=\frac{1}{2}\int_{0}^tV(0,t')dt' \nonumber \\
&=\frac{1}{\pi}t\arctan(1/t)+\frac{1}{2\pi}\ln(1+t^2) \label{eq:exsurfdisp}
\end{align}
The behavior of which as $t\rightarrow\infty$ is
\begin{equation}
u_z(0,0^+,t)= \frac{1}{\pi}+\frac{1}{\pi}\ln(t)
\label{eq:exsurfdispasy}
\end{equation}
which, apart from the constant of integration $u_0$, we could have deduced directly from (\ref{eq:surfdispasy}) given that $a_1=2$ for this example, and $y=0^+$ for the point of interest here. In Figure \ref{fig:exhalfspace} we overlay the asymptotic solution on the full displacement history for this example, and show the rapid convergence of the latter to the former.

\

To clearly outline the interplay between continuum deformation and fault rheology and to highlight its consistency with observations, we have thus far considered relatively simple model geometries; however, an extensive effort of using forward models of frictional faults embedded in a continuum to account for geodetic afterslip observations has accompanied both the increase in GPS data and computational resources needed to run many iterations of the forward models for parameter inversion. \emph{Linker and Rice} [1997] sought to reproduce the observed postseismic deformation following the Loma Prieta earthquake, considering several model fault geometries while comparing the response to a sudden step in stress of a linear viscoelastic fault rheology with a so-called hot-friction model, $\tau\sim\ln V$, intended to mimic the steady-state behavior of rate- and state-dependent friction. Both models provided comparable fits to the surficial displacements, with the conclusion that while deep relaxation process may be adequate, discriminating among different rheological models remains an issue. This continued to be reflected in subsequent efforts, including that of \emph{Hearn et al.} [2002], who considered geodetic observations of postseismic deformation in response to the 1999 Izmit earthquake. The authors performed a parameter inversion using a forward model to determine whether postseismic deformation in response to a co-seismic step in shear stress arises by viscoelastic relaxation of the lower crust, poroelastic rebound, or via a fault slip on regions at depth whose strength obeys a linear viscous strength of the type examined here or the logarithmic hot-friction model also used in \emph{Linker and Rice} [1997] and \emph{Marone et al.} [1991]. While poroelastic rebound was quickly ruled out and the frictional afterslip models offered improvements over the strictly viscoelastic model, the discrepancies between fault models with linear and logarithmic slip-rate dependence were marginal. 

\

That different frictional models may have consistent asymptotic behavior is further supported when considering numerical studies of stable afterslip. \emph{Hetland et al.} [2010] examined numerical solutions for the post-seismic response to imposed co-seismic stress changes on vertical strike-slip (anti-plane) faults that obeyed a wide range of frictional constitutive relations, ranging from linear and power-law non-linear viscous to logarithmic hot friction and rate- and state-dependence. The authors focus on the latter two relations and the numerical results show that the long-time decay of slip rates at a fixed position follow the expected $1/t$ decay anticipated by our analytical treatment here, given that the models examined are expected to relax back to the finite, model-driving slip rate about which these non-linear constitutive relations can be linearized. Our expected $1/t$ decay, as well as a spatial spreading of elevated slip rates that is proportional to time, is also reflected in several other numerical studies of rate-strengthening faults [e.g., \emph{Ariyoshi et al.}, 2007; \emph{Perfettini and Ampuero}, 2008]

\

While the issue of determining the relative contributions of different mechanisms to postseismic deformation remains, cases in which afterslip appears to be the dominant mechanism [e.g., \emph{Freed}, 2007] do appear to be well modeled as the accelerated creep of a fault with a rate- (and possibly state-) dependent friction [e.g., \emph{Johnson et al.}, 2006; \emph{Barbot et al., 2009}]; however, such studies  focus on a single fault strength description or rheology and comparatively little effort has been made to determine whether a particular rate-strengthening rheology is called for among several plausible ones [\emph{Mont\'esi}, 2004]. That disparate, non-linear rheological relations may share a common asymptotic behavior indicates that such an effort to discriminate requires careful attention. 


\subsection{Moment-duration scalings of spontaneous slow slip events}
\label{sec:momdur}

Another potential observational constraint on fault strength follows from slow slip events on subduction faults that occur without the accompaniment of a large earthquake. These events were first observed geodetically [\emph{Hirose et al.}, 1999; \emph{Dragert et al.}, 2001] and subsequently found to be accompanied by seismic tremor [\emph{Rogers and Dragert}, 2003; \emph{Obara et al.}, 2004; \emph{Obara and Hirose}, 2006). Compiled estimates of moment release and duration of subduction zone slow slip events appear to show a linear relationship between these quantities [\emph{Schwartz and Rokoksky}, 2007; \emph{Ide et al.}, 2007; \emph{Aguiar et al., 2009}; \emph{Gao et al.,} 2012; \emph{Liu}, 2014]. Slow slip events were shown to emerge spontaneously from fault models with slip rate- and state-dependent friction under marginal conditions for frictional instability [\emph{Liu and Rice}, 2005, 2007; \emph{Rubin}, 2008]. Numerical models are capable of producing a range of event sizes with a moment-duration relation that is arguably linear [\emph{Shibazaki et al.}, 2012; \emph{Liu}, 2014; \emph{Li and Liu}, 2016; \emph{Romanet et al.}, 2018]; however, a mechanistic explanation for the emergence of such a scaling has been missing, making it difficult to assess the robustness of and the necessary conditions for the observed model scaling.


\

Here we highlight alternative conditions permitting the existence of slow slip and show that a linear moment-duration scaling arises naturally from the stable response of a rate-strengthening fault to a stress perturbation or sudden elevation of pore fluid pressure. For simplicity, we focus on the example of a fault undergoing in-plane or anti-plane rupture, in which slip varies only along one dimension; however, the same conclusion can be reached following a similar line of argument given comparable slip-rate self-similarity on a fault undergoing mixed-mode rupture, in which slip varies in two dimensions. 

\

For a fault in which slip varies along one dimension, the moment per unit fault thickness in the out-of-plane direction is defined as
\begin{equation}
M(t)=\mu \int_{-\infty}^{\infty}\delta(x,t)dx
\end{equation}
and the moment rate is, using dot notation to denote a derivative with respect to time,
\begin{equation}
\dot M(t)=\mu \int_{-\infty}^{\infty}V(x,t)dx
\end{equation}
where in the two equations above we have momentarily returned to using dimensional variables. We now pass to non-dimensional variables as done to arrive to (\ref{eq:el}) and for the moment and moment rates using $M(t)/(\mu\delta_c x_c)\Rightarrow M(t)$ and $\dot M(t)/(\mu V_c x_c)\Rightarrow \dot M(t)$. We are interested in the asymptotic evolution of the fault moment release with time duration $t$ following a step in stress. Substituting the asymptotic expression for fault slip rate following a step in stress (\ref{eq:Vstepasym}), we find that the moment rate to leading order is constant, i.e.,
\begin{equation}
\dot M(t)= \frac{a_1}{\pi t}\int_{-\infty}^{\infty}\frac{1}{1+(x/t)^2}dx=a_1
\end{equation}
The constant $a_1$ can be interpreted as the net increase of force on the fault, per unit out-of-plane fault thickness, owed to the distributed stress step: i.e., for a stress step in the form $\tau_b(x,t)=\Delta\tau(x)H(t)$
\begin{equation}
a_1=\int_{-\infty}^\infty \Delta\tau(x)dx
\label{eq:cons}
\end{equation}
This representation is seen by recognizing that the integral defining $a_1$ 
\begin{equation}
a_1=\int_{-\infty}^\infty V(x,0)dx
\end{equation}
directly follows, given the relation $\tau=V$, from the more general statement of conservation for $t>0$
\begin{equation}
a_1=\int_{-\infty}^\infty \tau(x,t)dx
\end{equation}
which is itself deduced by demonstrating that, for $t>0$
\begin{equation}
\frac{d}{dt}\int_{-\infty}^\infty\tau(x,t)dx=0
\end{equation}
by substituting the expression (\ref{eq:el}) for the fault shear stress $\tau$, recognizing that $\partial [ \mathcal{H}(\partial\delta/\partial x)]/\partial t=\partial \mathcal{H}(V)/\partial x$, and using the condition that $\mathcal{H}(V)$ vanishes at $x=\pm\infty$. This implies that, for $t>0$
\begin{equation}
a_1=\int_{-\infty}^\infty \tau_b(x,t)dx
\end{equation}
from which (\ref{eq:cons}) follows. 

\

The above conservation condition on the net shear force exerted on the fault by the shear stress change, which implied a constant moment rate, is independent of the fault elastic configuration and mode of slip. In other words, while we considered the particular case of in- or anti-plane slip between elastic half-spaces, a constant moment rate would also be expected for any configuration---including mixed-mode slip between elastic half-spaces, or in proximity to a free surface---in which the fault strength is proportional to the sliding rate. By corollary, this also implies that a constant moment rate would ultimately be expected for non-linear, rate-dependent (and rate-strengthening) descriptions of fault strength, which can be considered to approach an effective linear-viscous response as the fault relaxes back to a steady sliding velocity. 

\

\begin{figure}[t]
\centering
 \noindent\includegraphics[scale=1.0]{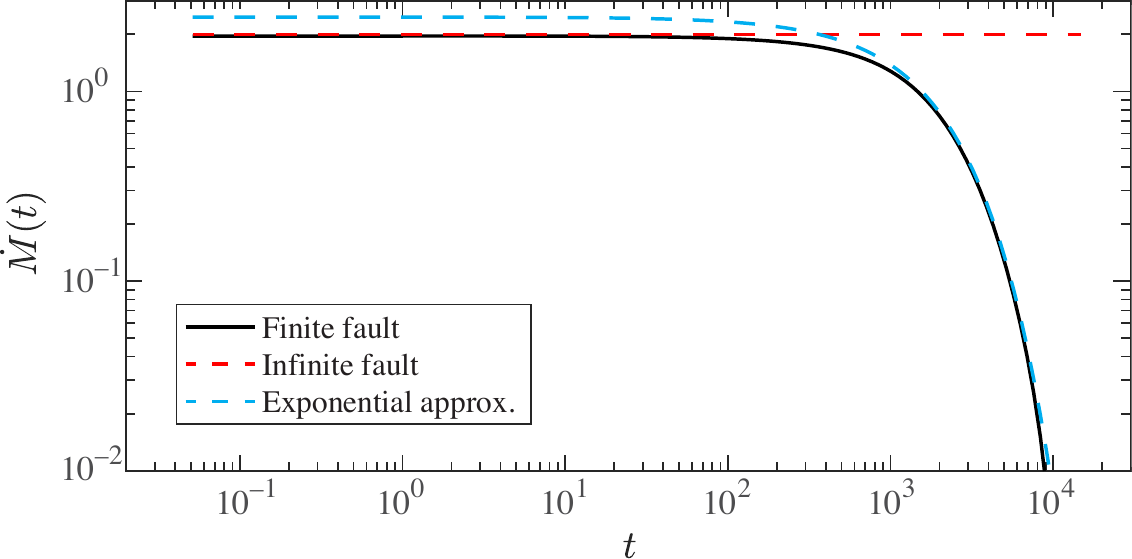}
 \caption{\small Plot of moment rate with time following a boxcar step in stress of unit magnitude and within a unit distance about the origin of a finite fault whose half-length $L=1000$. The initial moment rate is constant, as expected for an unbounded fault (red-dashed), followed by the exponential decay (cyan-dashed) expected from the finite-fault analysis.} \label{fig:FFMoment}
 \end{figure}

The duration of the constant moment rate cannot be indefinite. As noted in Section \ref{sec:FF}, locked boundaries will lead to exponential decay of the slip and moment rates. In Figure \ref{fig:FFMoment} we show this transition in the moment rate of a fault that experiences a sudden step in stress at $t=0$ with a boxcar distribution of unit magnitude, as in Sections \ref{sec:fullsoln} and \ref{sec:boxcarasy}, except here the fault is of finite length $2L=2000$, a length much larger than the initial step in stress occuring over distances $|x|<1$. As expected, at early times the finite fault boundaries are not apparent and the initial moment rate is constant and corresponds to the integrated step in stress, or equivalently the coefficient $a_1$, implying $\dot M=2$. At late times, the fault boundaries are manifest and the rapid, exponential decay of the moment rate follows. The exponential decay is captured by the dominant mode of the expansion (\ref{eq:FFsoln}), whose spatial distribution is approximated by (\ref{eq:omeg1}), the exponential decay rate is $\lambda_1=-1.157774...$, and the prefactor $v_1\approx0.0017$ is calculated using the initial condition via (\ref{eq:coeff}).

\

\

While we have thus far shown that the moment release following a step in stress is initially linear in time, we have not yet identified potential mechanisms for such a step that would initiate a slow slip event. While there may be potential causes for a fault to experience an increase in shear stress, we briefly outline here conditions by which an increase in pore fluid pressure could lead to a slow slip event with a linear moment-duration relationship. Specifically, we assume fault strength is frictional in nature and rate-strengthening with the simple form $f(V)=f_o+c V$, where here and in what follows we continue the dimensionless representation for brevity. The frictional nature of the strength implies that the shear strength can be written as $\tau_s=\bar\sigma f$ where $\bar\sigma$ is the effective normal stress $\bar\sigma=\sigma- p$, the difference between the total fault-normal stress $\sigma$ and the fault pore fluid pressure $p$. We consider a sudden increase in pore fluid pressure beyond an initial value $p_o$ of the form $p(x,t)=p_o+\Delta p(x)H(t)$, where we neglect here details of the fluid pressure evolution, but simply assume the rise time is short and the distribution is compact in space. The requirement that $\tau_s=\tau$ when and where sliding occurs can then be rearranged to have a form comparable to that of a shear stress step on a fault with linear viscous strength
\begin{equation}
\frac{\partial V}{\partial t}=\mathcal{H}\left(\frac{\partial V}{\partial x}\right)+\Delta\tau(x)\delta_D(t)
\end{equation}
for which we have just shown that the integral of $\Delta\tau(x)$ is a conserved quantity that determines the constant moment rate at long times. For the problem of a pore pressure step considered here, we may identify $\Delta\tau(x)$ with $f_o\Delta p(x)$. Therefore, such a sudden increase in pore pressure would also  give rise to the transient slip phenomena discussed in the preceding sections for local steps in shear stress.

\section{Conclusion}

We examined the evolution of slip and slip rate between elastic half-spaces that is accommodated as the shear of an adjoining thin, viscous layer
. The model provides an elementary description of stable, yet transient, fault slip and also provides a basis with which to examine non-linear, rate-dependent descriptions of fault strength. The problem reduces to an integro-differential equation for slip or slip rate that has an analogy with the classical diffusion equation, in which slip or slip rate takes the place of temperature and sources of fault shear stress take the place of sources of heat. The new problem governing fault slip evolution, which we refer to as Hilbert diffusion, has several distinctive features. In classical diffusion, the interaction among points in space is mediated by second-order spatial derivatives and is short-ranged, having a characteristic exponential decay; for viscous slip between elastic half-spaces, the interaction is now long-ranged with a characteristic power-law decay. Furthermore, while classical diffusion exhibits diffusive spreading proportional to the square-root of time, the Hilbert diffusion problem exhibits spreading proportionally to time, such that distinct features, like local maxima of slip or stress rates, may have an apparent propagation velocity.

\

The response of faults to sudden changes in shear stress yields a slip rate that decays as $1/t$ such that near-fault displacements grow logarithmically with time. Examining a simple fault system, this logarithmic growth of displacement extends to that observed at the surface. This suggests that the observed logarithmic time-dependence of geodetic, post-seismic displacement measurements may be a symptom of post-seismic slip on rate-strengthening faults. This is a relaxation from previous, post-seismic slip models of such deformation, which hinged on a specific non-linear constitutive form of fault rate strengthening (i.e., logarithmic dependence on slip rate) and the single-degree-of-freedom nature of the elastic deformation. Instead, we suggest that, when considering the continuum deformation of an elastic medium in response to fault slip, it may not be possible to discriminate on the particular form of the non-linearity of the frictional strength description on the basis of long-term surface measurements. This suggestion is provided that fault boundaries or rheological transitions do not play a determining role in limiting the spatial penetration of post-seismic slip. Furthermore, we highlight that, in the absence of locked boundaries, a linear rate-strengthening rheology necessarily implies that the scaling of moment with duration is linear in time. This scaling is consistent with inferences of such a relationship for slow slip transients and suggests that slow slip may be a manifestation of a rate-strengthening fault response to transient increases in fault shear stress or pore fluid pressure, the latter provided that the rate-dependence of strength is frictional in nature.

\appendix
\setcounter{equation}{0}
\renewcommand{\theequation}{A.\arabic{equation}}
\section{Solution for Green's function}
Here we provide details of the solution for the Green's function of the Hilbert diffusion equation. We denote the combined Fourier and Laplace transform of the slip distribution in space and time, respectively, as
\begin{equation}
D(k,s)=\int_{-\infty}^\infty e^{-2\pi i kx}\int_{-\infty}^\infty e^{-st} \delta(x,t)dt dx
\end{equation}
Taking the combined transform of
\begin{equation}
\frac{\partial\delta}{\partial t}=\mathcal{H}\left(\frac{\partial\delta}{\partial x}\right)+\delta_D(x-x')\delta_D(t-t')
\label{eq:Gprob}
\end{equation}
 and using the properties of each transform with respect to derivatives, convolutions, and the Dirac delta leads to 
\begin{equation}
sD(s,k)=-2\pi|k|D(k,s)+e^{-2\pi i kx' }e^{-st'}
\end{equation}
or, after rearranging,
\begin{equation}
D(s,k)=\displaystyle\frac{e^{-2\pi i kx' }e^{-st'}}{s+2\pi |k|}
\end{equation}

\

The inverse Laplace transform of $A(k)\exp(-st')/[s+B(k)]$ is $A(k)\exp[-B(k)(t-t')]H(t-t')$ where $A(k)=e^{-2\pi i kx'}$ and $B(k)=2\pi |k|$ and $H(x)$ is the Heaviside step function. The Green's function (7) then follows from the inverse Fourier transform. Ando et al. [2012], following Ida [1974], effectively arrived to this solution previously when considering a problem equivalent to (\ref{eq:Gprob}) with $x'=0$ and $t'=0$, though did not explicitly consider their solutions in terms of a fundamental solution as done here.

\section{Similarity solutions for slip rate}
\renewcommand{\theequation}{B.\arabic{equation}}
We look for similarity solutions for slip rate satisfying the Hilbert diffusion equation
\begin{equation}
\frac{\partial V}{\partial t} = \mathcal{H}\left(\frac{\partial V}{\partial x}\right)
\label{eq:VHilb}
\end{equation}
of the form $V(x,t)=U[\eta(x,t),\tau(t)]$ where $\eta=x/t$, $\tau=\ln t$ [e.g., \emph{Barenblatt}, 1996]. Substituting our ansatz into (\ref{eq:VHilb}),
\begin{equation}
\frac{\partial U}{\partial \tau}-\eta\frac{\partial U}{\partial \eta}=\mathcal{H}\left(\frac{\partial U}{\partial \eta}\right)
\end{equation}
We now look for solutions decomposed as
\begin{equation}
U(\eta,\tau)=f(\eta)\exp(\lambda \tau)
\end{equation}
which leads to the eigenvalue problem for eigenmodes $\lambda$ and eigenfunctions $f(\eta)$
\begin{equation}
\lambda f-\eta f'=\mathcal{H}(f')
\label{eq:simeq}
\end{equation}
If $f$ is an analytic function, we may write its real and imaginary parts as $f(\eta)=g(\eta)+ih(\eta)$, where $g=\mathcal{H}(h)$ (Appendix D). Consequently, $\mathcal{H}(f)=i f$, and (\ref{eq:simeq}) reduces to 
\begin{equation}
\lambda f =(i+\eta)f'
\end{equation}
which has the solution
\begin{equation}
f(\eta)=A(i+\eta)^\lambda
\end{equation}
where $A$ is a complex constant. Imposing boundary conditions that $f$ vanishes as $\eta\rightarrow\infty$, restricts $\lambda\leq 0$ and further requiring that $f$ have similar asymptotic behavior as $\eta\rightarrow\pm \infty$ implies that $\lambda$ has an integer value: i.e., $\lambda=-n$ where $n=1,2,3...$ and we denote the set of eigenfunctions as 
\begin{equation}
f_n(\eta)=\frac{i}{\pi(i+\eta)^n}
\label{eq:fndef}
\end{equation}
where the choice of prefactor $A=i/\pi$ is made for $\text{Re}[f_1(\eta)\exp(-\tau)]$ to correspond with the Green's function.

\section{Long-time asymptotic expansion of initial value problem}
\renewcommand{\theequation}{C.\arabic{equation}}
We are interested in determining the long-time asymptotic behavior of solutions for a sudden step in stress of the form 
\begin{equation}
\tau_b(x,t)=T(x)H(t)
\end{equation}
which can also be considered as an initial value problem 
\begin{equation}
V(x,0)=T(x)
\end{equation}
Specifically, we look to determine the complex coefficients $c_n=a_n\textcolor{red}{-}ib_n$ for an asymptotic expansion of slip rate of the form
\begin{equation}
V(x,t)=\sum_{n=1}^\infty \frac{1}{t^n} \text{Re}[c_n f_n(\eta)]
\label{eq:exp2}
\end{equation}
for an initial value problem in slip rate.

\

We begin with the Green's function solution to the problem
\begin{equation}
V(x,t)=\frac{1}{\pi t} \int_{-\infty}^\infty V(x',0)\frac{1}{1+\left[(x-x')/t\right]^2}dx'
\label{eq:asystart}
\end{equation}
Recalling the Taylor expansion about $\epsilon=0$
\begin{equation}
\frac{1}{1+\epsilon}=1-\epsilon+\epsilon^2-\epsilon^3+...=\sum_{k=0}^\infty (-1)^k \epsilon^k
\label{eq:Taylor}
\end{equation}
we may expand the Green's function about large time as 
\begin{equation}
V(x,t)=\frac{1}{\pi t} \int_{-\infty}^\infty V(x',0)\left[\sum_{k=0}^\infty (-1)^k\left(\frac{x-x'}{t}\right)^{2k}\right]dx'
\label{eq:interim}
\end{equation}
Using the binomial expansion
\begin{equation}
\left(a+b\right)^j=\sum_{i=1}^j\begin{pmatrix} j \\ i \end{pmatrix} a^{j-i}b^i
\end{equation}
we may rewrite (\ref{eq:interim}) as
\begin{equation}
V(x,t)=\frac{1}{\pi t} \int_{-\infty}^\infty V(x',0)\left[\sum_{k=0}^\infty\sum_{i=1}^{2k} \frac{(-1)^{k+i}}{t^{2k}} \begin{pmatrix} 2k \\ i \end{pmatrix} x^{2k-i}x'^{\,i}\right]dx'
\end{equation}
and swapping the order of summation
\begin{equation}
V(x,t)=\frac{1}{\pi t} \int_{-\infty}^\infty V(x',0)\left[\sum_{i=0}^\infty\sum_{k=0}^{\infty} \frac{(-1)^{k+i}}{t^{2k}} \begin{pmatrix} 2k \\ i \end{pmatrix}x^{2k-i}x'^{\,i}\right]dx'
\end{equation}
where the binomial coefficient is understood here to be zero if $2k<i$. We may then rearrange 
\begin{equation}
V(x,t)=\sum_{i=0}^\infty \frac{1}{t^{i+1}} \left[\int_{-\infty}^\infty V(x',0)x'^{\,i}dx'\right]\left[\frac{1}{\pi}\sum_{k=0}^{\infty} (-1)^{k+i} \begin{pmatrix} 2k \\ i \end{pmatrix}(x/t)^{2k-i}\right]
\label{eq:expsum}
\end{equation}
We identify the first term in brackets as the $i$-th moment of the initial distribution, which we defined as the coefficients $a_{i+1}$ in (\ref{eq:an}). We define the second term in brackets as the function $\rho_i(\eta)$, where $\eta=x/t$. For $i=0$, 
\begin{equation}
\rho_0(\eta)=\frac{1}{\pi}\sum_{k=0}^{\infty} (-1)^{k}(\eta)^{2k}=\frac{1}{\pi}\frac{1}{1+\eta^2}
\label{eq:rho0}
\end{equation}
where the latter equality follows from the Taylor expansion (\ref{eq:Taylor}). We note that this Taylor expansion suffers from a limited radius of convergence ($\eta^2<1$) owing to poles in the function being expanded. Such would not be an issue, for instance, when following a series of analogous steps to derive asymptotic expansions for the classical diffusion equation; however, we nonetheless can proceed here to derive the asymptotic expansion for Hilbert diffusion.

\

Given the definition of $\rho_{i}$, a recursion relation follows
\begin{equation}
\rho_{i+1}(\eta)=\frac{-1}{i+1}\rho_i'(\eta)
\end{equation}
Furthermore, since $\rho_0(\eta)$ is identically $\text{Re}[f_1(\eta)]$ and $f_n(\eta)$ follows the recursion relation $f_{n+1}(\eta)=-f'_n(\eta)/n$, we deduce that
\begin{equation}
\text{Re}[f_n(\eta)]=\rho_{n-1}(\eta) \quad\text{for } n=1,2,...
\end{equation}
such that the last bracketed term in (\ref{eq:expsum}) is seen to be the Taylor expansion of $\text{Re}[f_{i+1}(\eta)]$ and we may rewrite (\ref{eq:expsum}) as 
\begin{equation}
V(x,t)=\sum_{n=1}^\infty \frac{1}{t^n} a_n\text{Re}[f_n(\eta)]
\label{eq:Vasyreal}
\end{equation}

\

If the Hilbert transform of the initial distribution $V(x,0)$ exists, there will also be complementary terms to the asymptotic expansion (\ref{eq:Vasyreal})
\begin{equation}
V(x,t)=\sum_{n=1}^\infty \frac{1}{t^n} a_n\text{Re}[f_n(\eta)]+\sum_{n=1}^\infty \frac{1}{t^n} b_n\text{Im}[f_n(\eta)]
\label{eq:Vasyimag}
\end{equation}
To show the existence of the latter additional terms, we would begin by repeating the procedure that lead to (\ref{eq:Vasyreal}) but substituting $\mathcal{H}[V(x,t)]$ and $\mathcal{H}[V(x',0)]$ for $V(x,t)$ and $V(x',0)$ in (\ref{eq:asystart}), since $\mathcal{H}[V(x,t)]$ also satisfies the Hilbert diffusion equation and hence its solution can also be found by such a Green's function convolution. Doing so, we would arrive to an expression  similar to (\ref{eq:Vasyreal}) above
\begin{equation}
\mathcal{H}[V(x,t)]=\sum_{n=1}^\infty \frac{1}{t^n} b_n\text{Re}[f_n(\eta)]
\label{eq:asyend}
\end{equation}
where the coefficients $b_{i+1}$ are identified as the i-th moment of the initial distribution of  $\mathcal{H}[V(x',0)]$. Taking the inverse Hilbert transform of (\ref{eq:asyend}) we would arrive to the second term in (\ref{eq:Vasyimag}), after recalling that $\mathcal{H}(\text{Re}[f_n(\eta)])=-\text{Im}[f_n(\eta)]$. This accounts for the complete asymptotic expansion for slip rate as given in (\ref{eq:exp2}) and (\ref{eq:exp}).

\section{Relation between conjugate harmonic functions on the real line}

\renewcommand{\theequation}{D.\arabic{equation}}

The Cauchy integral formula for a complex analytic function $f$ of the complex variable $w$ states that, for a point $v$ on the contour $C$ [e.g., \emph{Carrier et al.}, 1983],

\begin{equation}
f(v)=-\frac{i}{\pi}\oint_C\frac{f(w)}{w-v}dw
\end{equation}

Take $C$ to be the semicircular contour of radius $R$ whose straight segment travels along the horizontal axis and whose arc lies in the upper half plane. For a function $f$ that vanishes far from the origin, we may take $R\rightarrow\infty$ such that the contour integral reduces to an integral along the real line and

\begin{equation}
f(\eta,\xi=0)=-\frac{i}{\pi}\int_{-\infty}^\infty\frac{f(s,r=0)}{s-\eta}ds
\end{equation}

where the complex numbers $v=\eta+i\xi$ and $w=s+i r$. Denote the real and imaginary parts of $f$ as

\begin{equation}
f(\eta,\xi)=g(\eta,\xi)+ih(\eta,\xi)
\end{equation}

where $g$ and $h$ are the conjugate harmonic functions satisfying the Cauchy-Riemann conditions.
Substituting (3) into (2),

\begin{equation}
g(\eta,0)+ih(\eta,0)=-\frac{i}{\pi}\int_{-\infty}^\infty\frac{g(s,0)+ih(s,0)}{s-\eta}ds
\end{equation}

Equating the real and imaginary parts of (4), we arrive to the relations

\begin{equation}
g=\mathcal{H}(h),\quad h=-\mathcal{H}(g)
\end{equation}

where we continue to use the shorthand $\mathcal{H}(f)=(1/\pi)\int_{-\infty}^\infty f(s)/(s-\eta)ds$, suppressing explicit reference to the null imaginary parts of $v$ and $w$.

\section*{Acknowledgments}
R. C. Viesca gratefully acknowledges support as a Professeur Invit\'e at MINES ParisTech, as well as from NSF grants EAR-1653382 and 1344993, and the Southern California Earthquake Center (this is SCEC contribution  8132). SCEC is funded by NSF Cooperative Agreement EAR-1033462 and USGS Cooperative Agreement G12AC20038. The presented results are readily reproduced following the detailed analytical methods.

\

\end{document}